\documentstyle[12pt]{article}

\begin{document}

\title{Why $E = mc^2$ Emerges in the Process of Neutron Capture}

\author{
{\large Ezzat G. Bakhoum}\\
\\
{\small Univ. of West Florida}\\
{\small 11000 Univ. Parkway, Pensacola, FL. 32514 USA}\\
{\small Email: bakhoum@modernphysics.org}\\
\\
{\small Copyright \copyright 2007 by Ezzat G. Bakhoum}\\
\\
\\
\begin{minipage}{6in}
\begin{center}
{\bf Abstract}
\end{center}
{\small 
This paper is a short commentary on the 2005 paper in Nature by S. Rainville et al., which claimed to be ``the most precise direct test of the famous equation", $E = mc^2$. This communication is directed only to the readers who are familiar with the earlier papers by the author on the subject of mass-energy equivalence.}
\end{minipage}
}

\date{}

\maketitle

In a paper published in Nature in December of 2005, entitled: ``A Direct Test of $E=mc^2$" \cite{Rainville}, author Simon Rainville et al. demonstrated the results of an experiment in which the fundamental process of neutron capture by nuclei of sulfur and silicon results in gamma radiation the energy content of which matches precisely the quantity $\Delta mc^2$. The authors claimed that the experiment represents ``the most precise direct test of the famous equation".\\
\\
It seems that the physics community has totally forgotten that Einstein made {\em two separate conclusions} concerning the concept of mass-energy equivalence in his famous paper of 1905. The first conclusion predicts that the kinetic energy of a particle is numerically equivalent to $\Delta mc^2$, and that if kinetic energy is converted to electromagnetic radiation then the energy content of that radiation would precisely match that quantity. That conclusion, fortunately, was correct (see the main paper by the author on this subject \cite{Bakhoum}). In the last few lines of his paper, however, Einstein made a second, ``general conclusion", which he gave without any proof (amazingly, after 100 years, that general conclusion is still without any proof): the conclusion that the {\em total energy} of a particle is equal to the total mass multiplied by $c^2$. Unfortunately, Einstein's ``general conclusion" \underline{was incorrect}! (see ref. \cite{Bakhoum}).\\
\\
What Rainville et al. have demonstrated in their experiment, as a matter of fact, is the equivalence between kinetic energy and electromagnetic energy (Einstein's first conclusion). The process of neutron capture by nuclei, unfortunately, has nothing to do with Einstein's ``general conclusion" (the direct conversion of mass into energy). That general conclusion can only be tested in particle decays. The reason is the following: in the process of neutron capture, a neutron is attracted to the nucleus by the strong nuclear force and becomes trapped inside a potential well \cite{Williams}. It is currently recognized that the partons (quarks) composing the neutron will lose part of their kinetic energies once inside the potential well \cite{Cramer}. This also means that the neutron will lose part of its apparent mass and actually become ``lighter" inside the nucleus (this accounts, for example, for the mass difference between the $^{29}$Si nucleus and its components, $^{28}$Si + neutron ). Now, the quarks' loss of kinetic energy does of course result in the emission of a gamma ray, which, thanks to the Rainville et al. experiment, is found to have an energy that matches precisely the quantity $\Delta mc^2$. Notice, however, that {\em no direct mass to energy transformation} ever occurs in the process of neutron capture. The Rainville et al. experiment therefore proves {\em only} Einstein's conclusion concerning kinetic energy.\\
\\
How about the plethora of experiments that have attempted in the past to prove Einstein's ``general conclusion", or the direct conversion of mass into energy? Except in the very few cases where the particle velocities actually approach the speed of light, they all failed! (see the previous papers by the author).\\
\\
{\large\bf \underline{For the Record}:}\\
\\
This communication was rejected by the journal Nature without any review. I am highly indebted to the arXiv, to the journal Physics Essays, and to other ``free speech" scientific publication media.

\end{document}